\documentstyle[12pt,aasms4]{article}

\def\etal{et\ al.}
\def\g{{$\gamma$}}

\received{1995 December 14}
\accepted{1996 February 14}

\lefthead{Quashnock}
\righthead{Constraint on the Distance Scale to GRBs}

\begin{document}

\title{A Constraint on the Distance Scale to\\
    Cosmological Gamma--Ray Bursts}

\author{Jean M. Quashnock\altaffilmark{1}}

\affil{Department of Astronomy and Astrophysics \\
        University of Chicago, Chicago, IL 60637}

\altaffiltext{1}{{\it Compton} GRO Fellow}

\begin{abstract}

If \g--ray bursts are cosmological in origin,
the sources are expected to trace the large--scale structure
of luminous matter in the universe.
I use a new likelihood method that compares the
counts--in--cells distribution of \g--ray bursts 
in the BATSE 3B catalog with that
expected from the known large--scale structure of the universe,
in order to place a constraint on the distance scale to cosmological bursts.
I find, at the 95\% confidence level, that the comoving distance to the
``edge'' of the burst distribution is greater 
than $630~h^{-1}$~Mpc ($z > 0.25$),
and that the nearest burst is farther than $40~h^{-1}$~Mpc.
The median distance to the nearest burst is $170~h^{-1}$~Mpc,
implying that the total energy released in \g--rays during a burst event 
is of order $3\times 10^{51}~h^{-2}$ ergs.
None of the bursts that have been observed by BATSE are in nearby galaxies,
nor is a signature from the Coma cluster or the ``Great Wall'' likely to be
seen in the data at present.

\end{abstract}


\section{Introduction}

The origin of \g--ray bursts is still unknown
and is currently the subject of a ``great debate''
in the astronomical community.
Do the bursts have a Galactic origin (\cite{L95}) or
are they cosmological (\cite{Pac95})?
And what is their distance scale?

In this Letter, I do not attempt to answer the first question,
but rather, I show that {\it if} one assumes that \g--ray bursts 
are cosmological in origin,
one can begin to answer the second question and place a
constraint on the distance scale to the bursts.
This is because cosmological bursts
are expected to trace the large--scale structure
of luminous matter in the universe (\cite{LQ93}, hereafter, LQ).
The constraint comes from
comparing the {\it expected} clustering pattern of bursts on the sky
--- which will depend on their distance scale because of projection effects ---
with that {\it actually observed}. 
The observed angular distribution is in fact quite isotropic (\cite{Brig96});
hence, only a lower limit to the distance scale can be placed
because a sufficiently large distance will always lead to 
a sufficiently isotropic distribution on the sky.

Hartmann and Blumenthal (1989) first used the absence of a
significant angular correlation function
in a small sample of \g--ray bursts 
to put a lower limit of $71~h^{-1}$~Mpc on the distance scale to
the bursts.\footnote{I follow the usual
convention and take $h$ to be the Hubble constant
in units of 100 km~s$^{-1}$~Mpc$^{-1}$.}
This lower limit was subsequently improved to $150~h^{-1}$~Mpc
(\cite{BHL94}) using the larger BATSE 1B catalog (\cite{Fish94}).

Here I use a powerful new likelihood method,
which I had previously developed to analyze repeating of
\g--ray bursts in the BATSE 1B and 2B catalogs (\cite{Q95}),
to compare the observed counts--in--cells distribution
in the new BATSE 3B catalog (\cite{Meeg96}) with that expected
for bursts at cosmological distances.
I describe this method
and calculate the expected counts--in--cells distribution in \S ~2.1,
using the angular correlation function computed in \S ~2.2.
I present my results on the burst distance scale in \S ~3,
and discuss some implications of this work in \S ~4.
This work was presented in preliminary form elsewhere (\cite{Q96}).

\section{Likelihood Method}

\subsection{Counts--in--Cells Distribution}

Let $N_{\rm cell}$ be a large number of circular cells,
each centered on a random position on the sky.
Each cell is of fixed solid--angle size $\Omega=2\pi(1-\cos\theta_{\rm rad})$,
where $\theta_{\rm rad}$ is the angular radius of the cell.
I set the number of cells to be such that any part of the sky is covered,
on average, by one cell; hence, $N_{\rm cell}=4\pi/\Omega$.
Let $C_N$ to be the number of these cells having $N$ \g--ray bursts in them,
out of the $N_{\rm tot}=1122$ in the BATSE 3B catalog,
where $N=0,1,2,\ldots$.
Then I define the observed counts--in--cells distribution,
\begin{equation}
P_N\equiv C_N/N_{\rm cell}\; , 
\end{equation}
which is the probability that a randomly chosen cell of size $\Omega$
has $N$ bursts in it.
The counts--in--cells distribution contains information about clustering
of \g--ray bursts on scales comparable 
to the angular size $\theta_{\rm rad}$ of the cell.  

I now define $Q_N$ to be the counts--in--cells distribution that is expected
if \g--ray bursts are cosmological in origin and 
trace the large--scale structure of luminous
matter in the universe.
This expected distribution depends on only one unknown parameter,
the effective distance $D$ to \g--ray bursts (which I define below),
because the angular clustering pattern of bursts on the sky will depend
by projection on this distance.
 
The {\it likelihood} ${\cal L}$
measures how likely it is that the observed counts--in--cells 
distribution $P_N$ is drawn from the expected distribution $Q_N$.
Since $Q_N$ depends on the unknown effective distance $D$ to \g--ray bursts,
the likelihood is really a measure of how likely a given value of $D$ is.
It is given by the multinomial expression
\begin{equation}
{\cal L} = {N_{\rm cell}!\over \prod_N C_N!} \prod_N Q_N^{C_N}\; .
\end{equation}
Combining equations (1) and (2), and noting that the factorial term is
parameter independent, I obtain
\begin{equation}
\log{\cal L} = N_{\rm cell} \sum_N P_N \log Q_N + {\rm constant}\; .
\end{equation}

The expected counts--in--cells distribution $Q_N$ is most easily computed
using the generating function ${\cal Q}(\lambda)\equiv \sum_N Q_N\lambda^N$
(\cite{Peeb80}; \cite{BS89})
and writing it in terms of the means,
over the cell size,
of the irreducible $N$--point angular
correlation functions $\bar w_N$ (\cite{Wh79}; \cite{BS89}):
\begin{equation}
{\cal Q}(\lambda)={\rm exp}\left[\langle N\rangle(\lambda -1) +
\sum_{N=2}^\infty \bar w_N {\langle N\rangle^N\over N!}
(\lambda -1)^N \right] \; ,
\end{equation}
where the density 
$\langle N\rangle = N_{\rm tot}/N_{\rm cell}$ 
is the mean number of bursts in a cell.
Indeed, the Poisson distribution, 
$Q_N = \langle N\rangle^N {\rm exp}\left(-\langle N\rangle\right)/N!$,
is obtained (by comparing powers of $\lambda$) 
when all correlations $\bar w_N$ are zero 
(uniform distribution) and only the first term in the sum 
in equation (4) is included.

Now for cosmological \g--ray bursts, the clustering is expected to be weak (LQ)
and can be adequately described by one number, $\bar w_2$,
the mean of the two--point angular correlation function 
$w(\theta)$ over the cell:
\begin{equation}
\bar w_2 = {1\over \Omega^2} 
\int_\Omega d\Omega_1\,d\Omega_2\, w(\theta_{12}) \; .
\end{equation}
Then only the first two terms in the sum in equation (4) contribute 
to the generating function,
and the counts--in--cells distribution is a convolution of two terms:
$Q_N = \sum_{M=0}^N A_{N-M}\, B_M$, 
where the first is a Poisson uncorrelated background
of single bursts of density $\langle N\rangle -\langle N\rangle^2 \bar w_2$,
and the second a uniform smattering of correlated pairs, 
with density $\langle N\rangle^2 \bar w_2/2$,
in which
\begin{eqnarray}
A_N & = & {1\over N!} (\langle N\rangle -\langle N\rangle^2 \bar w_2)^N
e^{-(\langle N\rangle -\langle N\rangle^2 \bar w_2)}\; ,
\nonumber \\ & & \\
B_{2k} & = & {1\over k!} (\langle N\rangle^2 \bar w_2/2)^k 
e^{-\langle N\rangle^2 \bar w_2/2}\; .\nonumber
\end{eqnarray}

Once the angular correlation function is known, 
$\bar w_2$ can be computed from equation (5),
and the expected counts--in--cells distribution 
$Q_N$ can be found from equation (6).
The likelihood ${\cal L}$ can then be computed from equation (3),
and it is a function
of the unknown effective distance $D$ through 
the dependence of $\bar w_2$ on $D$.

\subsection{Angular Correlation Function}

For simplicity, I assume that $\Omega_0=1$ and $\Lambda=0$,
and that the large--scale structure clustering pattern is constant in comoving
coordinates.  The results (cf. \S ~3) are, in fact, 
insensitive to these assumptions because of the
small redshifts that are involved.

The Limber equation (\cite{Peeb80}), 
which relates the angular correlation function 
$w(\theta)$ to the spatial one,
can be recast (\cite{P91}; LQ)
in terms of the dimensionless power spectrum $\Delta^2(k)$:
\begin{equation}
w(\theta)={\int_0^\infty r^4 \phi^2(r)\,dr 
\int_0^\infty \pi \Delta^2(k) J_0(kr\theta)\,dk/k^2 \over
\left[\int_0^\infty r^2 \phi(r)\,dr\right]^2}\; .
\end{equation}

Here the selection function $\phi(r)$ is the
probability that a source at comoving distance $r$ produces 
a burst that is in the BATSE catalog, i.e.,
a burst with apparent flux greater than
the limiting flux of the BATSE survey.
The selection function is a convolution of a source function,
representing the luminosity function and the spatial density of the sources,
with an observer function, representing the efficiency with which BATSE
detects bursts of a given flux.

Now the cumulative $C_{\rm max}/C_{\rm min}$ distribution
of \g--ray bursts seen by BATSE begins to roll over from a $- 3/2$ power law
for bursts fainter than $C_{\rm max}/C_{\rm min} \sim 10$ (\cite{Meeg92}).
Since this is many times above threshold,
it suggests that BATSE sees most of the source distribution
and that this distribution is not spatially homogeneous.
To the extent that this is the case, 
the observer function is unity for nearly
all the burst sources, irrespective of brightness, 
so that BATSE sees nearly all of the burst sources.
The selection function $\phi(r)$, then,
depends only on the luminosity function and
the mean space density of the sources.

I define $D$ as the distance beyond which 
$\phi(r)$ drops appreciably; 
thus, $D$ is the {\it effective distance}
to the ``edge'' of the source distribution in the BATSE catalog. 
One could approximate the selection function to be unity 
out to a comoving distance $D$, and
zero beyond, but evolution of the luminosity function 
and/or the mean space density of the
sources will change the form of $\phi(r)$ from $\theta(r-D)$;
more generally, I define the effective distance $D$ 
as ${1\over 3} D^3 \equiv \int r^2\,\phi(r)\, dr$ (LQ). 
Thus, $D$ is not the distance to the very dimmest burst in the BATSE catalog, 
but rather the typical distance to most of the dim bursts in the sample.

Equation (7) then becomes, upon rescaling $y\equiv r/D$,
\begin{equation}
w(\theta)= {9\pi\over D} \int_0^1 y^4\, dy 
\int_0^\infty \Delta^2(k)\, J_0(kyD\theta)\, dk/k^2 \; ,
\end{equation}
so that the angular correlation function
satisfies the well--known scaling relation 
(\cite{Peeb80}) $w(\theta)= D^{-1}W(D\theta)$ as a function of $D$.

I take the power spectrum that characterizes the large--scale
clustering of \g--ray burst sources 
to be the same as that determined from a redshift 
survey of radio galaxies (\cite{PN91}):
\begin{equation}
\Delta^2(k)={0.129\, (k/k_0)^4\over 1 + (k/k_0)^{2.4}} \; ,
\end{equation}
where $k_0=0.025\, h\; {\rm Mpc}^{-1}$.
This power spectrum is characteristic of moderately rich environments,
and is intermediate between that of ordinary galaxies and clusters (LQ).
Because the exact bias factor relating the clustering of \g--ray burst sources 
to that of luminous matter is unknown,
such an intermediate {\it Ansatz} is reasonable.
In any case, the resultant distance limit depends only weakly
on the bias factor (roughly as the square root).

Substituting equation (9) into equation (8), one obtains the expected
intrinsic angular correlation function $w(\theta)$ 
of \g--ray bursts as a function of their
effective depth $D$. 
Figure 1 shows $w(\theta)$ ({\it dashed lines}) for $D$ of 500, 1000, and
2000 $h^{-1}$~Mpc. Note the scaling of $w(\theta)$ as a function of $D$.

Now the observed angular correlation function, $\widetilde w(\theta)$,
is smeared at small angular scales because of finite positional errors.
Each burst in the BATSE catalog is assigned 
a positional uncertainty $\theta_{\rm err}$
corresponding to a $68\%$ confidence that the true burst position
is within an angle $\theta_{\rm err}$
to the position listed in the catalog.
Defining $\sigma^2\equiv \left(1-\cos(\theta_{\rm err})\right)/1.14$,
the smearing function can be approximated as a Gaussian of the angular
separation $\Delta\theta$ between the true and observed positions of the burst:
$dP/d\Omega ={\rm exp}(-\Delta\theta^2/2\sigma^2)/2\pi\sigma^2$.
If all burst positions are smeared by the same amount, $\sigma^2$,
the observed angular correlation function is a convolution of the
intrinsic correlation function with the smearing function and the
modified Bessel function $I_0$ (\cite{HLB91}):
\begin{equation}
\widetilde w(\theta) = \int {\phi\,d\phi\over 2\sigma^2}\,w(\phi)\,
e^{-(\theta^2+\phi^2)/4\sigma^2}
\,I_0\left({\theta\phi\over 2\sigma^2}\right) \; .
\end{equation}

Figure 1 shows $\widetilde w(\theta)$ ({\it solid lines}) 
for $D$ of 500, 1000, and 2000 $h^{-1}$~Mpc, 
with smearing of burst positions of $\theta_{\rm err}=3\fdg 8$,
the median value in the BATSE 3B catalog.
Note the smearing of the correlation function
on scales smaller than $\theta_{\rm err}$,
along with the aliasing of some small--scale power to angular scales
comparable to the positional smearing.
It is this smeared correlation function that I use
in equation (5) when calculating $\bar w_2$ and the expected
counts--in--cells distribution $Q_N$,
in calculating the likelihood $\cal L$ as a function of $D$.

The cell size
$\theta_{\rm rad}$ is chosen in order to maximize the sensitivity of detection,
or signal--to--noise ratio,
given the strength of the signal expected.
The signal $S$ is given by the total number of correlated pairs in the cells
(cf. above eq. [6]): $S=N_{\rm cell}\langle N\rangle^2\bar w_2/2$.
The noise $N$ is the square root of the total number of pairs
(\cite{Peeb80}; LQ): $N=\left(N_{\rm cell}\langle N\rangle^2/2\right)^{1/2}$.
For a sample of 1122 bursts (corresponding to the total number of bursts in the
BATSE 3B catalog) with positional smearing of $\theta_{\rm err}=3\fdg 8$,
the signal--to--noise ratio is maximized when 
cells of $\theta_{\rm rad}=5^\circ$ are used.

\section{Results}

Figure 2 shows the likelihood
of the BATSE 3B catalog data as a function
of the effective comoving distance $D$,
calculated using cells of size $\theta_{\rm rad}=5^\circ$.
The likelihood is normalized to that expected for an isotropic
distribution on the sky.
At large values of $D$ 
(the maximum value allowed is $D=R_{\rm H}=6000~h^{-1}$ Mpc,
the size of the horizon in a closed universe),
the likelihood goes to unity, because
by projection a sufficiently large distance will always lead to an isotropic
distribution on the sky. Note also that there is no value of $D$ for which
the likelihood is greater than 1; 
thus, the maximum likelihood value for $D$ is $R_{\rm H}$,
and the 3B data are consistent with isotropy.

The solid line in Figure 2
shows the likelihood for a positional smearing of $\theta_{\rm err}=3\fdg 8$,
corresponding to the median value in the 3B catalog.
To illustrate the dependence of these results on
positional errors, I also show ({\it dashed line}) the results 
for a larger positional smearing\footnote{Graziani \& Lamb (1996) 
compare the 3B positions with those
from the IPN network 
and conclude that the systematic errors are larger than the $1\fdg 6$
value quoted in the 3B catalog. 
Their best--fit model gives a median positional error of $6\fdg 6$.}  
of $\theta_{\rm err}=6\fdg 6$
(with cells of size $\theta_{\rm rad}=9^\circ$ 
to maximize the signal--to--noise ratio).

Small values of the effective comoving distance to
\g--ray bursts are unlikely, according to Figure 2;
I find, at the 95\%
confidence level, that for the 3B median positional error
of $\theta_{\rm err}=3\fdg 8$,
$D$ must be greater than $630~h^{-1}$ Mpc,
corresponding to a redshift $z > 0.25$.
If the positional errors are larger than quoted 
and are better characterized
by $\theta_{\rm err}=6\fdg 6$, these results are only slightly
weakened;
at the 95\% confidence level,
$D$ must be greater than $500~h^{-1}$ Mpc,
corresponding to a redshift $z > 0.19$.

These limits are not sensitive to earlier
assumptions (\S ~2.2) on cosmology and clustering
evolution, since these only become important at
higher redshifts.
They are also conservative limits, in that a constant median
value for the positional errors was used rather than the entire
distribution of errors.
This is because the bright bursts, which ostensibly are
nearer to us, are more clustered (by the scaling property of
the correlation function, eq. [8]) and are responsible for
the bulk of the expected signal, but, in fact, have smaller
errors than the median value. The faint bursts, which
are far away, are hardly clustered to begin with
(even before smearing), but have errors larger than the median value.
Hence, the expected clustering pattern has been smeared more
by using a constant median value (this permits a simpler calculation
and allows eq. [10] to be used) than by smearing using
the entire distribution of errors.
Therefore, the counts--in--cells statistic has been weakened somewhat,
and thus the quoted lower limits are, in fact, conservative.

\section{Discussion}

If \g--ray bursts are cosmological and trace the large--scale
structure of luminous matter in the universe,
and their  positional errors are as quoted in the 3B catalog,
then the lack of any angular clustering in the data
implies that the observed distance to the ``edge'' of the  burst distribution
must be farther than $630~h^{-1}$ Mpc.
Since there are 1122 bursts in the catalog, 
an effective limit on the {\it nearest} burst to us can be
placed by convoluting the likelihood as a function of $D$ (Fig. 2) with
the nearest neighbor distribution of 1122 bursts inside a sphere of
radius $D$. I find that the nearest burst
must be farther than $40~h^{-1}$ Mpc at the 95\% confidence level,
and farther than $10~h^{-1}$ Mpc at the 99.9\% level.
At this level of confidence, then, none of the bursts that have been
observed by BATSE are in nearby galaxies.
A signature from the Coma cluster or the ``Great Wall'' 
($\sim 70~h^{-1}$ Mpc) is not likely to be
seen in the data at present, since only a few bursts
could have originated from these distances.
Indeed, a search for such a signature (\cite{HBM96})
found no compelling evidence for anisotropy in
supergalactic coordinates.

The median distance to the nearest burst is $170~h^{-1}$ Mpc.
Since the brightest burst in the 3B catalog
has a fluence of $7.8\times 10^{-4}~{\rm ergs~cm}^{-2}$ in \g--rays,
this implies that the total energy released in \g--rays during a burst event 
is of order $3\times 10^{51}~h^{-2}$ ergs.

As the number of observed \g--ray bursts keeps increasing,
the distance limit will improve.
In fact, LQ showed that, with 3000 burst
locations, the clustering of bursts might just be detectable
and would provide compelling evidence for a cosmological origin.
If it is not detected, the redshift to the ``edge'' of the bursts
would be put at $z\sim 1$ or beyond.

\acknowledgments

I would like to acknowledge useful discussions with Carlo Graziani,
Don Lamb, Cole Miller, and Bob Nichol.
This research was supported in part by NASA through the {\it Compton}
Fellowship Program --- grant NAG 5-2660, grant NAG 5-2868,
and contract NASW-4690.

\clearpage

\begin{figure}
\plotone{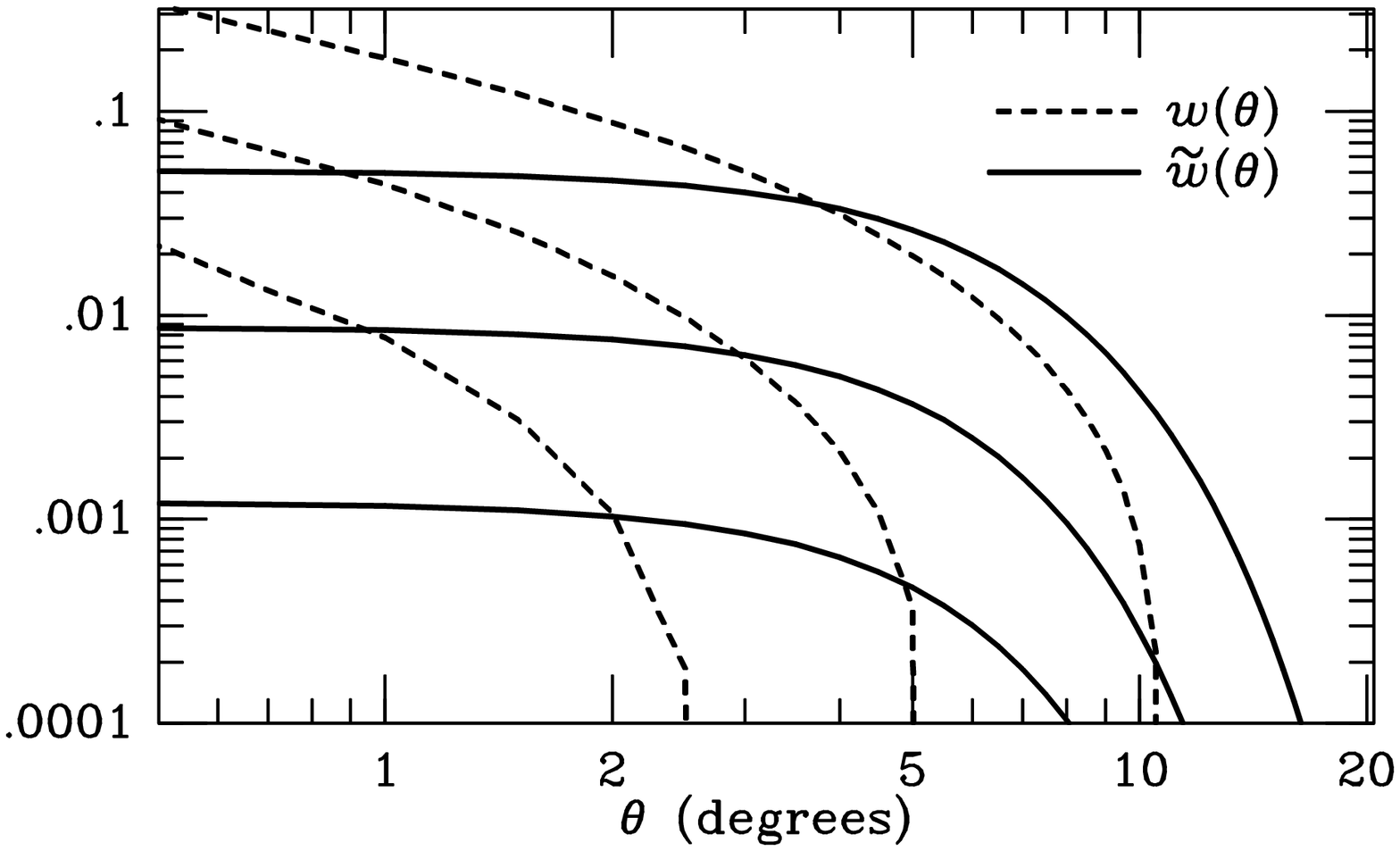}
\figcaption[fig1.eps]{Angular correlation function and its dependence on the
effective comoving distance $D$ to \g--ray bursts. 
Shown are both the intrinsic,
$w(\theta)$ ({\it dashed line}), and smeared, $\widetilde w(\theta)$ 
({\it solid line}), correlation functions, in decreasing amplitude,
for $D$ = 500, 1000, and 2000 $h^{-1}$~Mpc.
The smearing corresponds to $\theta_{\rm err}=3\fdg 8$,
the median value in the BATSE 3B catalog (see text).}
\end{figure}

\begin{figure}
\plotone{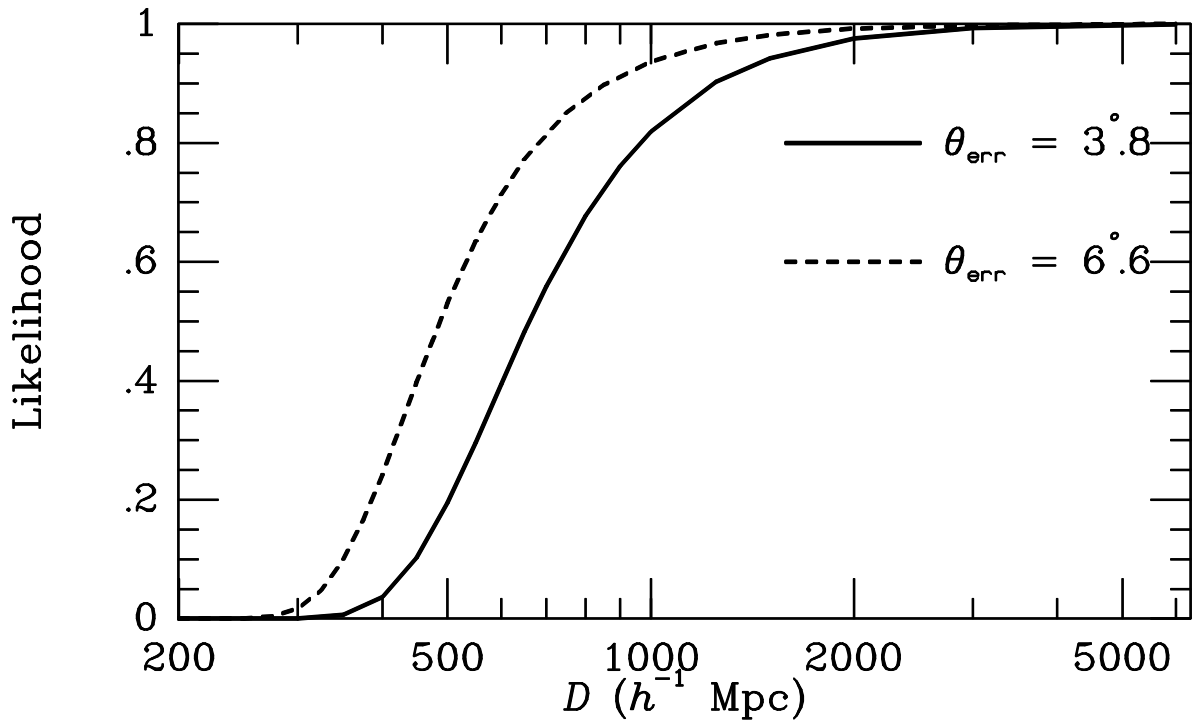}
\figcaption[fig2.eps]{Likelihood of the BATSE 3B catalog data as a function
of the effective comoving distance $D$ to \g--ray bursts,
shown with a smearing of $\theta_{\rm err}=3\fdg 8$ and $6\fdg 6$ (see text).}
\end{figure}

\end{document}